\documentclass[conference]{IEEEtran}
\IEEEoverridecommandlockouts
\usepackage{cite}
\usepackage{amsmath,amssymb,amsfonts}
\usepackage{algorithmic}
\usepackage{graphicx}
\usepackage{textcomp}
\usepackage[table]{xcolor}
\usepackage{multirow}
\usepackage{multicol}
\usepackage{bm}
\usepackage{booktabs}
\usepackage{amssymb}% http://ctan.org/pkg/amssymb
\usepackage{pifont}% http://ctan.org/pkg/pifont

\usepackage{utfsym}

\usepackage{wasysym}
\newcommand{\cmarks}{\checked}
\usepackage{adjustbox}

\def\BibTeX{{\rm B\kern-.05em{\sc i\kern-.025em b}\kern-.08em
    T\kern-.1667em\lower.7ex\hbox{E}\kern-.125emX}}
\begin{document}

\title{Multi-Level Speaker Representation for Target Speaker Extraction
\thanks{$\dag$ Correspondence to Shuai Wang (wangshuai@cuhk.edu.cn)} 
}
\author{\IEEEauthorblockN{{Ke Zhang$^{1, 2}$, Junjie Li$^{3}$, Shuai Wang$^{2,\dag}$, Yangjie Wei$^{1}$, Yi Wang$^1$, Yannan Wang$^4$, Haizhou Li$^{2,5,6}$}}
\IEEEauthorblockA{$^1$Key Laboratory of Intelligent Computing in Medical Image, Northeastern University, China}
\IEEEauthorblockA{$^2$ SDS, SRIBD, The Chinese University of Hong Kong, Shenzhen, China}
\IEEEauthorblockA{$^3$ The Hong Kong Polytechnic University, Hong Kong, China} 
\IEEEauthorblockA{$^4$Tencent Ethereal Audio Lab, Tencent, Shenzhen, China}
\IEEEauthorblockA{$^5$Machine Listening Lab (MLL), University of Bremen, Germany}
\IEEEauthorblockA{$^6$Department of Electrical and Computer Engineering, National University of Singapore, Singapore}
}

\maketitle

\begin{abstract}
Target speaker extraction (TSE) relies on a reference cue of the target to extract the target speech from a speech mixture. While a speaker embedding is commonly used as the reference cue, such embedding pre-trained with a large number of speakers may suffer from confusion of speaker identity. In this work, we propose a multi-level speaker representation approach, from raw features to neural embeddings, to serve as the speaker reference cue. We generate a spectral-level representation from the enrollment magnitude spectrogram as a raw, low-level feature, which significantly improves the model's generalization capability. Additionally, we propose a contextual embedding feature based on cross-attention mechanisms that integrate frame-level embeddings from a pre-trained speaker encoder. By incorporating speaker features across multiple levels, we significantly enhance the performance of the TSE model. Our approach achieves a 2.74 dB improvement and a 4.94\% increase in extraction accuracy on Libri2mix test set over the baseline.
\end{abstract}


\section{Introduction}
Target speaker extraction (TSE) aims at extracting the target speaker's speech from a mixture, which typically relies on a reference cue of the target, such as voiceprint~\cite{wang2018voicefilter, Liu_X_Sepformer, kamo23_interspeech,chen23k_interspeech, borsdorf2021universal,  sato22b_interspeech, Qian2023, Curriculum_interspeech}, facial information~\cite{pan2021reentry,usev21, li2024audio,li23ja_interspeech}, or even body movements~\cite{pan2022seg}. 
Audio-only TSE relies on a pre-recorded speech to extract the target speech. Therefore, an appropriate speaker feature representation for the target speaker becomes crucial for accurate target speech extraction. Failing to extract appropriate speaker features may lead to unexpected TSE outputs, such as broken utterances or incorrect speaker output, that is called speaker confusion problem~\cite{Liu_X_Sepformer,zhao22b_interspeech}.

The popular TSE models are jointly trained with an auxiliary module which generates the speaker embedding of the target speaker~\cite{ge2020spex+, delcroix2020improving, zmolikova21_interspeech}. 
However, such jointly trained speaker encoder lacks sufficient discrimination ability between  speakers\cite{zhang23k_interspeech} as far as TSE is concerned. 
Zhao et al.~\cite{zhao22b_interspeech} attribute the speaker confusion problem to the high acoustic similarity and limitation of speaker encoder for speaker characterization. To overcome the speaker confusion problem, Mu et al.~\cite{AAAI_2024} propose a self-supervised disentangled representation learning to remove the local semantic information considered irrelevant to speaker identity.

\begin{figure*}[ht]
  \centering
  \includegraphics[width=1\textwidth]{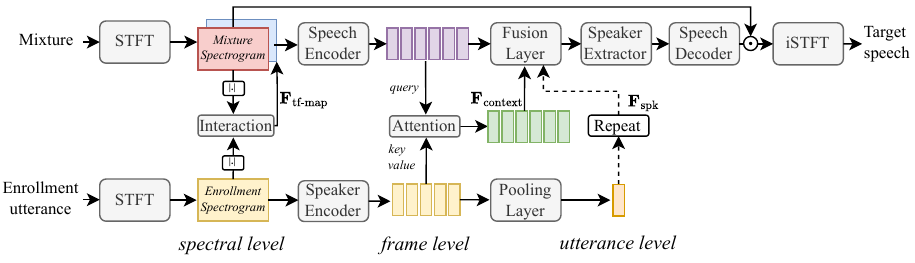}
  \vspace{-3mm}
  \caption{The architecture of the proposed target speaker extraction model with multi-level speaker features as the reference cue. The model consists of two main pipelines. The upper left-to-right pipeline is the speaker extraction module, while the lower left-to-right pipeline is a speaker encoder that extracts multi-level speaker representation. The lower pipeline serves as the reference cue of the upper pipeline in a target speaker extraction task. $\left| \cdot \right|$ denotes the magnitude operation. $\mathbf{F}_{\text{tf-map}}$, $\mathbf{F}_{\text{context}}$, and $\mathbf{F}_{\text{spk}}$ denote the TF Map feature, Contextual Embedding feature, and Speaker Embedding feature, respectively.
  % \textcolor{red}{(To Ke: the color in Fig 1 and 2 of the same module should be consistent. Please do not assign random color to the figures.!)} 
  } 
  \label{fig:Structure}
  \vspace{-5mm}
\end{figure*}

Different from the joint training method, early TSE models employ pre-trained speaker models trained for the speaker verification task to provide speaker embedding~\cite{wang2018voicefilter}. 
Although the speaker encoder trained on numerous speakers can produce embedding with high speaker discrimination, such speaker embedding does not perform satisfactorily on the TSE task due to bad generalization~\cite{quantitative}. 
Liu et al.~\cite{Liu_X_Sepformer, liu2023improving} attempt to solely train speaker encoders on TSE data, achieving better performance on in-domain datasets, while the encoder still exhibits limited speaker discrimination, raising concerns about the cross-domain effectiveness. 

Recent studies started to explore the use of frame-level embedding generated by speaker encoders, revealing that not only speaker identity information but also contextual information from the enrolled speech, despite different linguistic content, is beneficial for the TSE task~\cite{ContrastiveLearn, zhang23r_interspeech}. Moreover, some even skipped the use of speech encoders entirely, opting to fuse speaker information directly at the spectrogram level~\cite{ ExploitContext, SpecMatch}. The success of these methods suggests that raw, low-level speech information, without explicit extraction of speaker identity, seems to be sufficient for effective speaker extraction. Additionally, there has been a growing interest in leveraging large self-supervised speech models to construct TSE models~\cite{nttSelf, LinSPL}. The speaker features derived from these large self-supervised learning models have also contributed to improved TSE performance.

To investigate the practical contributions of speaker features at different granularity for TSE, we categorize speaker features into three levels based on the extent of speaker information extraction, from raw features to neural features. These levels include low-level spectral features, frame-level embeddings extracted by the speaker encoder, and utterance-level speaker embeddings derived to represent speaker identity. For low-level spectral features, we introduce a method to generate a representation from the enrollment magnitude spectrogram, referred to as the TF Map feature. The Contextual Embedding feature, which integrates raw contextual information with frame-level embeddings, is generated using a cross-attention mechanism that produces time-varying speaker embeddings for fusion. Finally, utterance-level neural speaker embedding serves as the high-level feature.

This paper is organized as follows. In Section II, we discuss the details of multi-level speaker features. In Section III, we report the experiment setup. In Section IV, we summarize the results. Finally, Section V concludes the discussion.
%In this work, we propose and validate the effectiveness of hierarchical speaker information modeling in the TSE task. Our findings indicate that low-level features can significantly enhance the performance of pre-trained speaker encoders in TSE tasks while also reducing the risk of overfitting. Experimental results show that, when using features from a single level, low-level features outperform higher-level features that primarily focus on extracting target speaker identity. This superior performance is largely due to the enhanced generalization capabilities of low-level features. Moreover, the combined use of multiple feature types leads to substantial improvements in the extraction performance and accuracy of the TSE model. These results highlight the critical importance of leveraging speaker information across different levels to optimize TSE performance.

\section{Proposed Method}

We discuss three-level speaker features in conjunction with a typical speaker extraction pipeline. The Band-Split RNN (BSRNN)~\cite{BSRNN} model is selected as the backbone for the speech extractor, and the pre-trained Ecapa-TDNN \cite{desplanques2020ecapa} is employed as the speaker encoder for extracting the speaker features. Fig.~\ref{fig:Structure} is the diagram of an overall architecture.

\begin{figure} 
  \centering
  \includegraphics[width=0.45\textwidth]{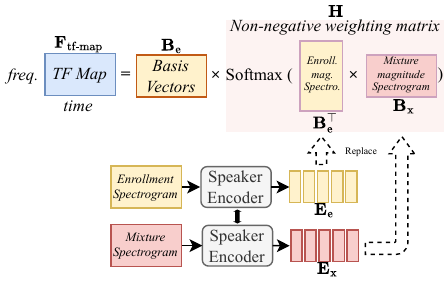}
  \vspace{-2mm}
  \caption{The calculation process of the TF Map feature. It consists of two non-negative components: basis vectors from the enrollment's magnitude spectrogram and a weighting matrix, computed based on either 1) Spectral Similarity or 2) Embedding Similarity between the mixture and the enrollment.}
  \label{figure: TFMap}
  \vspace{-4mm}
\end{figure}

\subsection{TF Map Feature} 
% \textcolor{red}{(to Ke: please elaborate the details in the captions of all Figures and Tables. Please also explain why we use TF Map Feature - you mention that you are inspired by XXX, it sounds copying from others without thinking. please explain our own thoughts. )}

Since the TSE model identifies the target speaker by learning the relationship between auxiliary information and the target speech component, establishing a strong correlation between the two is essential. Spectral components from the same speaker tend to exhibit similarities, such as shared spectral structures and harmonic content\cite{kinnunen2003spectral, hershey2016deep}, forming a natural connection between the enrollment and target speech within the mixture. This may provide an advantage over relying solely on high-level features like speaker embedding. Therefore, we investigate whether well-constructed spectral features can serve as effective cues for guiding the TSE model to extract the target speech.

% \junjie{Since spectral components from the same speaker naturally exhibit similarity, including similar spectral structures and harmonic content \cite{kinnunen2003spectral, hershey2016deep}, this similarity is preserved in the connection between the enrollment and target speech within the mixture. This suggests that using only the spectral information from the enrollment speech to reconstruct the target speech could be a possible approach. }

% \sout{Since the TSE task focuses on extracting the target speaker's voice rather than determining their presence in the mixture, we believe that establishing the correlation between the auxiliary information and the target speech takes priority over differentiating the auxiliary features from different speakers.
% Spectral components from the same speaker naturally exhibit similarity, including similar spectral structures and harmonic content. This similarity is preserved as a natural connection between the enrollment and target components in the mixture, offering an advantage that high-level features like speaker embeddings do not possess. Therefore, we explore whether constructing appropriate spectral features can serve as effective cues for target speech extraction.}

Inspired by the non-negative matrix factorization (NMF)~\cite{NMF, NMF_Conv, NMF_sparse} for speech reconstruction, we propose a method to generate a spectral-level representation based on the enrollment magnitude spectrogram, referred to as the TF Map feature. The generated speaker feature at the spectral level is concatenated with the spectrogram of the mixture during the extraction process, enhancing the target speech extraction.

The NMF method decomposes the magnitude spectrogram of clean speech to obtain non-negative basis vectors for individual speakers. These basis vectors are then iteratively combined through non-negative weighted summation to reconstruct the target speech in the mixture. Typically, the number of basis vectors is kept smaller than the feature dimensions, allowing the basis vectors to capture the intrinsic underlying spectral characteristics while minimizing correlation with interfering speech.

In this work, the basis vectors are simplified to include all frames of the magnitude spectrogram from the pre-enrolled utterance, and the non-negative weighting matrix is directly calculated based on the similarity using an attention mechanism, which is generated through linear correlation and a \textit{softmax} operation. Fig.~\ref{figure: TFMap} illustrates the workflow of the TF Map feature. 
% \textcolor{red}{(the symbols in Eq(1)(2) need to appear in Fig 2. as well.-Haizhou)}

\begin{equation}
    \label{eqa: tfmap}
    \mathbf{F}_{\text{tf-map}} =  \mathbf{B}_{\rm e} \mathbf{H}
\end{equation}
where $\textbf{B}_{\rm e} \in \mathbb{R}^{D_{\rm f}\times T_{\rm e}}$ denotes the magnitude spectrogram of the enrollment, and $\textbf{H} \in \mathbb{R}^{T_{\rm e} \times T_{\rm x}}$ denotes the non-negative weighting matrix. $D_{\rm f}$ denotes the dimension of frequency, $T_{\rm e}$ and $T_{\rm x}$ denote the frame length of enrollment and mixture.

Here we provide two approaches for calculating the non-negative weighting matrix based on the similarity between the mixture and enrollment:

\subsubsection{Spectral Similarity}
The similarity between spectral frames is used to compute the non-negative weighting matrix, and then normalized using a Softmax function:
\begin{equation}
    \label{eqa: score1}
    \mathbf{H} =  \text{Softmax} \bm{(\mathbf{B}_{\rm e}^\top \mathbf{B}_{\rm x}} )
\end{equation}
where $\textbf{B}_{\rm x} \in \mathbb{R}^{D_{\rm f}\times T_{\rm x}}$ denotes the magnitude spectrogram of the mixture. $^\top$ denotes the transpose operation.

\subsubsection{Embedding Similarity}
To more accurately assess the similarity between the mixture's components and the basis vectors from the enrollment, both the mixture and enrollment are transformed into the embedding space through the speaker encoder:
\begin{equation}
    \label{eqa: score2}
    \mathbf{H} =  \text{Softmax} \bm{(\mathbf{E}_{\rm e}^\top \mathbf{E}_{\rm x}} )
\end{equation}
where $\textbf{E}_{\rm e} \in \mathbb{R}^{D_{\rm e}\times T_{\rm e}}$ and $\textbf{E}_{\rm x} \in \mathbb{R}^{D_{\rm e}\times T_{\rm x}}$ denote the sequence of the frame-level embeddings from the enrollment and the mixture, respectively. $D_{\rm e}$ denotes the feature dimension of the frame-level embedding.

All vectors are length-normalized when employed in calculating the weighting matrix, and the cosine similarity is used for the comparison. 
% \textcolor{red}{to Ke: what do you mean here? Following the reconstruction of the target speaker's components, each frame in the TF Map feature is normalized by the projection of the mixture amplitude spectra.} 
Besides, the energy of the constructed TF Map feature is recovered by projecting the amplitude spectrogram of the mixed signal onto the corresponding frame of the feature.

\subsection{Contextual Embedding Feature}
The speaker encoder further projects the enrollment spectrogram into an embedding space to characterize the speaker. 
% \junjie{Here, we refer to the embedding sequences generated before the pooling layer as frame-level embeddings, preserving the variability along the time axis.}

Frame-level embeddings can be viewed as speaker embeddings with certain contextual information, retaining the length and variability along the time axis.  This characteristic provides significant flexibility when applied to TSE tasks. 
% \junjie{Because the length of frame-level embeddings can vary, we use a cross-attention mechanism to generate a Contextual Embedding feature that matches the length of the speech mixture representation. This approach allows for the adaptive integration of speaker information: }
In this work, we employ a cross-attention mechanism to generate a time-varying speaker feature at the embedding level, referred to as Contextual Embedding. This feature maintains the same length as the spectral frames of the mixture, enabling the adaptive integration of speaker information:
\begin{equation}
    \label{eqa: attention}
    \mathbf{F}_{\text{context}} =  \text{Softmax} \left(\frac{\mathbf{Q} \mathbf{K}^\top}{\sqrt{d_k}}\right) \mathbf{V}
\end{equation}
\begin{equation}
    \label{eqa: Q}
    \mathbf Q = \mathbf X^\top \mathbf{W^{\rm Q}}
\end{equation}
\begin{equation}
    \label{eqa: K}
    \mathbf K = \mathbf E_{\rm e}^\top \mathbf W^{\rm K}
\end{equation}
\begin{equation}
    \label{eqa: V}
    \mathbf V = \mathbf E_{\rm e}^\top \mathbf W^{\rm V}
\end{equation}
where $\mathbf{X} \in \mathbb{R}^{D_{\rm x}\times T_{\rm x}}$ is the 
% \junjie{speech mixture representation.} 
encoded mixture speech.
$D_{\rm x}$ is the feature dimension.
% the encoded mixture.
$\mathbf{W^{\rm Q}} \in \mathbb{R}^{D_{\rm x}\times d_k}$, $\mathbf{W^{\rm K}} \in \mathbb{R}^{D_{\rm e}\times d_k}$, and $\mathbf{W^{\rm V}} \in \mathbb{R}^{D_{\rm e}\times d_k}$ are learnable parameters for Query, Key and Value in attention module, respectively. $d_k$ is the dimension in the attention module.

\subsection{Speaker Embedding Feature}
Utterance-level speaker embedding, the most common and widely used form of speaker information, is typically extracted by a speaker encoder and a pooling layer, which is primarily focused on capturing the target speaker's identity information. Due to the lack of temporal variability, the speaker embedding is replicated along the time axis to align with the speech mixture representation, resulting in the Speaker Embedding feature 
$\mathbf{F}_{\text{spk}}$.
% the Speaker Embedding feature is usually generated by replicating across all frames of the mixed signal to achieve consistent speaker information fusion, denoted as $\mathbf{F}_{\text{spk}}$. 

In this work, we employ a multiplicative approach to construct the speaker information fusion layer, facilitating the integration of speaker identity information across the temporal dimension.
% frames. ~\cite{ochiai2019multimodal}

\section{Experimental Setup}
\label{section: Experiments}
\subsection{Dataset}
We conduct experiments using clean Libri2mix~\cite{libri2mix} dataset, consisting of mixtures of two speakers. 
The TSE models are trained with 251 speakers on Libri2mix-100 containing 13900 utterances. The validation and test set each contains 3000 utterances and 40 non-overlapping speakers. Every sample is used twice to extract two different speakers by switching the enrollment. Fully overlap (min. version) and 16 kHz sampling rate are configured.

\subsection{Training Details}
The BSRNN~\cite{BSRNN} is employed as the speech extractor in the TSE model. In the speech encoder, we split the frequency band below 1.5 kHz by a 100 Hz bandwidth, the frequency band between 1.5 kHz and 3.5 kHz by a 200 Hz bandwidth, the frequency band between 3.5 kHz and 6 kHz by a 500 Hz bandwidth, treat the frequency band between 6 kHz and 8 kHz as one subband. This split results in 32 subbands. The feature dimension of each subband is 128. The speaker extractor is constructed with a 192-dimensional bidirectional LSTM repeated 6 times in the BSRNN.

The speaker encoder utilizes the Ecapa-TDNN architecture and is pre-trained with 5994 speakers on VoxCeleb2~\cite{Voxceleb2}. The model is available in the open-source project WeSpeaker~\cite{wespeaker}\footnote{https://github.com/wenet-e2e/wespeaker}.

We train all TSE models for 150 epochs on segments with 3 seconds. Adam optimizer is used. The learning rate is set to $1e^{-3}$ and decays to $2.5e^{-5}$. Negative scale-invariant signal-to-noise ratio is selected as the loss function.

All the codes are implemented in PyTorch and will be released in the open source project WeSep~\cite{wesep} 
\footnote{https://github.com/wenet-e2e/wesep}.

\subsection{Evaluation Metrics}
The extraction performance is evaluated through the well-known metric: the SI-SDR improvement (SI-SDRi)~\cite{le2019sdr}. Besides, we consider the samples with SI-SDRi larger than 1 dB as the successful extraction. The percentage of these samples is measured for the accuracy of extraction.

% \begin{table}[htbp]
%     \centering
%     \vspace{-2mm}
%     \caption{The results of BSRNN with different speaker features.}
%     \addtolength{\tabcolsep}{-2pt}
%     \begin{tabular}{c|c|c|c|c|c|c}
%     \midrule
%     \multirow{2}*{Model} &\multicolumn{2}{c|}{TF Map} &Contextual &Speaker &SI-SDRi &Acc. \\ 
%      &Spec. &Emb. &Embedding &Embedding &/ dB &/ \% \\
%     \midrule
%     BSRNN &\cmark & & & &15.18 &96.60 \\
%     & &\cmark& & &15.25 &96.83 \\
%         & & &\cmark & &14.51 &95.27 \\
%     \cellcolor{gray!31}(baseline) &\cellcolor{gray!31} &\cellcolor{gray!31} &\cellcolor{gray!31} & \cellcolor{gray!31}\cmark &\cellcolor{gray!31}13.17 &\cellcolor{gray!31}92.08 \\
        
%     \cmidrule{2-7}
%         &\cmark & &\cmark & &15.52 &96.40 \\
%         &\cmark & & &\cmark &15.40 &96.07 \\
%     \cmidrule{2-7}
%        & &\cmark &\cmark & &\textbf{15.91} &\textbf{97.02} \\
%         & & \cmark & &\cmark &15.60 &96.63 \\
%     \midrule
%     \multicolumn{7}{l}{ ``Spec." and ``Emb." denote \textit{Spectra} and \textit{Embedding Similarity}, respectively.}\\
%     \multicolumn{7}{l}{ Acc. is the percentage of samples with SI-SDRi $>$ 1 dB.}
%     \end{tabular}
%     \vspace{-5mm}
%     \addtolength{\tabcolsep}{2pt}
%     \label{table: feature}
% \end{table}

\begin{table}[htbp]
    \centering
    \caption{The results of BSRNN with different speaker features.}
    \addtolength{\tabcolsep}{-3pt}
    \begin{tabular}{c|c|c|c|c|c}
    \midrule
    \multirow{2}*{Model} &\multirow{2}*{TF Map} &Contextual &Speaker &SI-SDRi &Accuracy \\
     & &Embedding &Embedding &/ dB &/ \% \\
    \midrule
    BSRNN & Spec.  & & &15.18 &96.60 \\
    &Emb. & & &15.25 &96.83 \\
        & &\cmarks & &14.51 &95.27 \\
    \cellcolor{gray!31}(baseline) &\cellcolor{gray!31} &\cellcolor{gray!31} &\cellcolor{gray!31}\cmarks &\cellcolor{gray!31}13.17 &\cellcolor{gray!31}92.08 \\
    \cmidrule{2-6}
        &Spec.&\cmarks & &15.52 &96.40 \\
        &Spec. & &\cmarks &15.40 &96.07 \\
    \cmidrule{2-6}
        &Emb. &\cmarks & &\textbf{15.91} &\textbf{97.02} \\
        &Emb. & &\cmarks &15.60 &96.63 \\
    \cmidrule{2-6}
        &Emb. &\cmarks &\cmarks &15.85 &96.95 \\
    \midrule
    % \multicolumn{6}{l}{ ``\cmarkd" denotes the similarity calculated in embedding space for TF Map.}\\
    \multicolumn{6}{l}{ ``Spec." and ``Emb." denote Spectral and Embedding Similarity, respectively.}\\
    \multicolumn{6}{l}{ Accuracy is the percentage of samples with SI-SDRi $>$ 1 dB.}
    \end{tabular}
    \vspace{-5mm}
    \addtolength{\tabcolsep}{2pt}
    \label{table: feature}
\end{table}

\section{Results and Discussion}
\label{section: results}

\subsection{Comparative Studies with Different Speaker Features}
Table~\ref{table: feature} shows the performance of the BSRNN with different speaker features. The model with only the speaker embedding as a speaker feature is selected as the baseline.  

Among the three speaker features, the TF-map feature achieves the best performance, followed by the Contextual Embedding. The Speaker Embedding feature performs the worst, with a 2.08 dB decrease in SI-SDRi and a 4.75 percentage point drop in accuracy compared to the best one.
% Within the TF-map features, the method that uses the embedding space to compute the weighting matrix yields the best performance. 

The use of multiple speaker features can significantly enhance the performance of TSE models. Models that incorporate multiple speaker features outperform those using a single feature. The best configuration combines both TF Map (Emb.) and Contextual Embedding features, leading to an improvement of approximately 0.7 dB in SI-SDRi and a 0.4\% increase in accuracy compared to the best single-feature configuration. These findings highlight that leveraging information from different levels within the enrollment utterance is crucial for optimizing the performance of TSE models. 

An interesting finding is that incorporating all three features yields performance comparable to that of using only the TF Map and Contextual Embedding features. We hypothesize that this is because the Contextual Embedding already captures the information contained in the Speaker Embedding feature.

% \begin{table}
%     \centering
%     \caption{SI-SDRi with a single feature on training and validation set.
%     }
%     \begin{small}
%     \vspace*{-2mm}
%     \addtolength{\tabcolsep}{-3pt}
%     \begin{tabular}{c|c|c|c} 
%        \toprule
%         Speaker Feature  &Training Set   &Validation Set &Degradation \\  
%         \midrule
%         TF Map (Spec.)  &17.14  &15.45   &1.69      \\
%         TF Map (Emb.) &17.19 &15.75 &1.44    \\
%         Contextual Embedding   &17.02  &14.63 &2.39         \\
%         Speaker Embedding    &17.56  &12.95 &4.61          \\ 
%        \bottomrule
%     \end{tabular}
%     \vspace*{-5mm}
%     \addtolength{\tabcolsep}{3pt}
%     \label{table:loss}
%     \end{small}
% \end{table}

\subsection{Generalization of Different Features}
% The observation that higher-level speaker features can worsen extraction performance may seem counterintuitive. Fig. \ref{fig:loss} illustrates the performance variations across the training, validation, and test sets, which may help explain this phenomenon.
% Higher-level speaker features show a greater performance gap between the training and test sets, indicating poor generalization. These features, such as the Speaker Embedding feature, primarily capture speaker identity, which encompasses a narrow range of speech information. This limits the data diversity available to the speaker extraction module during training. In contrast, lower-level features contain more comprehensive information beyond speaker identity, which can be implicitly utilized by the speaker extraction module, improving performance.

\begin{figure}[htbp]
    \centering
    \vspace{-3mm}
    \includegraphics[width=0.9\linewidth]{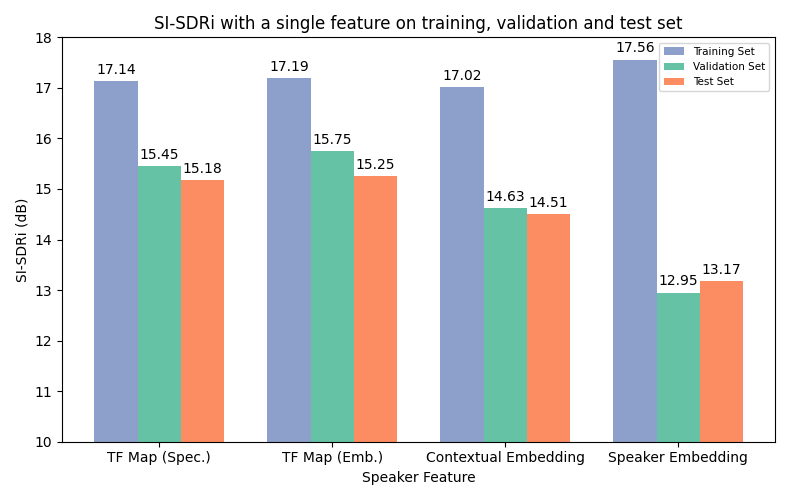}
    \vspace{-2mm}
    \caption{SI-SDRi with a single feature on training, validation, and test set.}
    \label{fig:loss}
    % \vspace{-1mm}
\end{figure}

The observation that higher-level speaker features can worsen extraction performance may seem counterintuitive. Figure \ref{fig:loss} illustrates performance variations across the training, validation, and test sets, which may help explain this phenomenon. Higher-level features, such as the Speaker Embedding, show a larger performance gap between the training and test sets, indicating poorer generalization. These features primarily capture speaker identity, which provides a limited range of speech information and restricts data diversity available to the speaker extraction module during training. In contrast, lower-level features offer more comprehensive information beyond speaker identity, allowing the extraction module to leverage a broader data spectrum, thus enhancing performance.

\begin{table}[htbp]
    \centering
    \vspace{-3mm}
    \caption{Comparison with other methods on Libri2mix. }
    \vspace{-1mm}
    \begin{adjustbox}{center}
    \addtolength{\tabcolsep}{-5pt}
    \begin{tabular}{c|c|c|c|c|c}
    \midrule
    \multirow{2}*{Model} &Speaker &Training &SI-SDRi &Acc. &\multirow{2}*{Pub.} \\
      &Model &Method &/ dB &/ \% &\\
    \midrule
    SpeakerBeam~\cite{delcroix2020improving} &ResNet &Joint &13.03 &95.2 &ICASSP, 2020 \\
    SpEx+$^{\dag}$~\cite{ge2020spex+} &ResNet &Joint &13.41 &- &Interspeech, 2020 \\
    sDPCCN~\cite{han2022dpccn} &ConvNet &Joint &11.61 &- &ICASSP, 2022 \\
    Target-Conf.$^{\dag}$~\cite{zhao22b_interspeech} &ResNet &Joint &13.88 &- &Interspeech, 2022 \\
    MC-SpEx$^{\dag}$~\cite{chen23k_interspeech} &ResNet &Joint &14.61 &- &Interspeech, 2023 \\
    X-T-TasNet~\cite{heo24_interspeech} &d-vector &Pretrained &13.48 &95.3 &Interspeech, 2024\\
    \midrule
    SSL-TD- &ResNet + &Pretrained &14.01 &96.1 &\multirow{2}*{ICASSP, 2024} \\
    SpeakerBeam~\cite{nttSelf} &WavLM &Finetuning &14.65 &97.0 & \\
    \midrule
    \multirow{2}*{BSRNN~\cite{LinSPL}} &Campplus + &\multirow{2}*{Pretrained} &\multirow{2}*{15.39} &\multirow{2}*{-} &\multirow{2}*{SPL, 2024} \\
        &SHuBERT & & & & \\
    \midrule
    BSRNN &\multirow{2.5}{*}{Ecapa-TDNN} &\multirow{2.5}{*}{Pretrained} &\textbf{15.91} &\textbf{97.0} &\multirow{2.5}*{Proposed} \\
    \cmidrule{1-1} \cmidrule{4-5}
    BSRNN\(^*\) & & &\textbf{17.99} &\textbf{98.6} & \\
    \midrule
    \multicolumn{6}{l}{ $^{\dag}$: trained under an 8kHz sampling rate.} \\
    \multicolumn{6}{l}{ \(^*\): trained on the Libri2mix-360.} \\
    \end{tabular}
    \addtolength{\tabcolsep}{5pt}
    \end{adjustbox}
    \vspace{-3mm}
    \label{table: compare}
\end{table}

\subsection{Comparison with Other Methods}

Table~\ref{table: compare} compares our results with other TSE models on the Libri2mix dataset. Recent studies \cite{LinSPL,nttSelf} explored the extraction of speaker features using large models trained through self-supervised learning (SSL). These approaches leverage not only pre-trained speaker encoders but also SSL-based speech models to extract speaker information. With the vast data and powerful encoding capabilities of SSL models, TSE models can utilize embeddings at various levels, tailored to the specific needs of the task, resulting in enhanced performance.

In contrast, our proposed method relies solely on the pre-trained ECAPA-TDNN model, providing a distinct advantage in terms of overall model size. Despite this, the strong performance of our approach shows that effectively utilizing information at different levels of the enrollment utterance alone can achieve competitive results. 

\section{Conclusion}
We validate the effectiveness of three levels of speaker features in target speaker extraction. We confirm that integrating low-level features with speaker identity information significantly improves performance. This approach enhances generalization and mitigates overfitting, resulting in more accurate extraction outcomes. 
% In future work, we will extend this research to noisy and reverberant environments to further assess the model's robustness.

% \section*{Acknowledgment}

% The preferred spelling of the word ``acknowledgment'' in America is without 
% an ``e'' after the ``g''. Avoid the stilted expression ``one of us (R. B. 
% G.) thanks $\ldots$''. Instead, try ``R. B. G. thanks$\ldots$''. Put sponsor 
% acknowledgments in the unnumbered footnote on the first page.

% \section*{References}

% Please number citations consecutively within brackets \cite{b1}. The 
% sentence punctuation follows the bracket \cite{b2}. Refer simply to the reference 
% number, as in \cite{b3}---do not use ``Ref. \cite{b3}'' or ``reference \cite{b3}'' except at 
% the beginning of a sentence: ``Reference \cite{b3} was the first $\ldots$''

% Number footnotes separately in superscripts. Place the actual footnote at 
% the bottom of the column in which it was cited. Do not put footnotes in the 
% abstract or reference list. Use letters for table footnotes.

% Unless there are six authors or more give all authors' names; do not use 
% ``et al.''. Papers that have not been published, even if they have been 
% submitted for publication, should be cited as ``unpublished'' \cite{b4}. Papers 
% that have been accepted for publication should be cited as ``in press'' \cite{b5}. 
% Capitalize only the first word in a paper title, except for proper nouns and 
% element symbols.

% For papers published in translation journals, please give the English 
% citation first, followed by the original foreign-language citation \cite{b6}.
\bibliographystyle{IEEEtran}
\bibliography{mybibliography}

\end{document}